\documentclass[12pt,oneside]{article}
\usepackage{amsmath}
\def\sect #1{\setcounter{equation}{0}}

\begin{document}
\renewcommand{\baselinestretch}{1.24} 

\title{\bf
Black holes vs. naked singularities formation in
collapsing Einstein's clusters}

\author{S. Jhingan\thanks{E-mail : sanjhi@mate.polimi.it} and G.
Magli\thanks{E-mail : magli@mate.polimi.it}\\
{\it Dipartimento di Matematica, Politecnico di Milano}\\
{\it Piazza Leonardo da Vinci, 32, 20133 Milano, Italy}}
\maketitle

\begin{abstract}
Non-static, spherically symmetric clusters of counter-rotating particles, of 
the type first introduced by Einstein, are analysed here. The initial data 
space can be parameterized in terms of three arbitrary functions, namely; 
initial density, velocity and angular momentum profiles. The final state 
of collapse, black hole or naked singularity, turns out to depend on the 
order of the first non-vanishing derivatives of such functions at the centre. 
The work extends recent results by Harada, Iguchi and Nakao.

\end{abstract}

\newpage

\section{Introduction}

The ``final fate'' of  gravitational collapse of spherically symmetric dust
clouds is well understood. The central singularity can be naked or covered 
depending on the choice of initial data; there is a ``critical branch'' of 
solutions where a transition from naked singularities to 
black holes occurs (see \cite{Global, Jhingan97} and references therein).

Recently, a considerable effort has been paid in order to 
extend the above mentioned results to collapse models with
general matter fields. In particular, one would like to understand the role
played by stresses in the formation and the nature of the
singularities, since both numerical~\cite{OriPiran, Harada1, rrs}
and analytical~\cite{Cooperstock, JD} studies tend to confirm violations of the
cosmic censorship conjecture of the kind observed in dust spacetimes. 

Gravitational collapse with stresses is quite difficult to handle analytically
even in spherical symmetry, one of the reasons being the lack of physically
valid exact solutions.
In a recent series of papers~\cite{Magli01, Magli02} we have obtained the 
general exact solution for the case of anisotropic materials
sustained  only by tangential stresses (see~\cite{Herrera} for a review on the
role of anisotropic materials in relativistic astrophysics);
the nature of the singularities forming in such solutions is still
largely unknown (exceptions are the self-similar case~\cite{Magli02}
and models with special equation of state~\cite{TPSingh, Barve}).

Among the solutions with vanishing radial stresses, there is a system
of counter-rotating particles - the ``Einstein cluster'' -
first introduced by Einstein~\cite{Einstein} (see also~\cite{Florides}) 
in the static case and then generalized to the non-static case by 
Datta~\cite{Datta} and Bondi~\cite{Bondi} (see also~\cite{he-s} and~\cite{ev}).
The motion of the particles in the cluster is sustained by angular momentum
whose average effect is to introduce a non-vanishing tangential pressure
in the energy-momentum tensor. 
This model is particularly interesting from the physical point
of view since it can give some hint on the effects of rotation on
collapse, without raising difficulties usually associated with the
axisymmetric spacetimes. The analysis of the singularities of the Einstein 
cluster model has been recently pioneered
by Harada, Iguchi and Nakao (~\cite{Harada}, to be referred afterwards as HIN).
Using the ``mass-area'' coordinate technique, developed in~\cite{Magli02},
these authors discovered a particular solution within the Einstein cluster 
class for which the quadratures can be explicitly carried out.
This allowed them to discuss the nature of the singularity
for this special solution in full details. 
They have also discovered several features of the singularities
forming in the general model, for which the metric is not
expressible in terms of elementary functions.
Their results comprise several new examples of naked singularities.

Our results here extend the HIN results 
to non-analytic cases.  We show that the initial data can be parameterized in 
terms of the first non-vanishing derivatives near the centre, as in the dust 
case, and obtain the corresponding spectrum of endstates of the collapse for 
marginally bound initial configurations.  Several new interesting features 
of the Einstein cluster model arise.

\section{The Einstein cluster model} 

The general solution of the Einstein field equations for
spherically symmetric collapse with tangential stresses can be
reduced to quadratures using a technique first introduced by 
Ori~\cite{Ori} to study the spherically symmetric charged dust.
We will very briefly recall the solution here, referring the reader to 
paper~\cite{Magli02} for details.

Systems with vanishing radial stresses are characterized by conservation of the
Misner-Sharp mass ($m$). Therefore, one can use it as a comoving label and 
introduce a system of coordinates (``mass-area coordinates'')
in which the line element turns out to be 
\begin{equation}
ds^2 = 
-K^2\left(1-\frac{2m}{R}\right)
dm^2+2
\frac{KE}{uh} 
dRdm - \frac 1{u^2} dR^2 + R^2 (d\theta^2 + \sin^2\theta d\varphi^2) \ .
\label{m01}
\end{equation}
Here $K$, $u$ and $h$
are functions of $R$ and $m$.
The function $u$ is the modulus of the velocity 
of the collapsing shells and satisfies 
\begin{equation}
u^2 =-1 + \frac{2m}{R} + \frac{E^2}{h^2} \ ,
\label{j}
\end{equation}
while
\begin{equation}
K = g(m) + \int G(m,R)  dR \ ,
\label{m10}
\end{equation}
where 
\begin{equation}
G(m,R) := \frac{h}{RE} \left[1+\frac{R}{2}\left(\frac{E^2}{h^2}\right)_{,m}
\right]\left(-1 + \frac{2m}{R} + \frac{E^2}{h^2} \right)^{-3/2} \ .
\label{m11}
\end{equation}
The functions $g(m)$ and $E(m)$ are arbitrary. The quantity $h(m,R)$ is a 
measure of the internal energy stored in the material, and therefore 
plays the role of equation of state. It can be shown 
that the quantities characterizing a material with vanishing
radial stresses, namely the energy density $\epsilon$ 
and  the tangential stress $\Pi$, 
can be written in terms of $h$ as follows~\cite{Magli01}:
\begin{equation}
\epsilon = \frac{h}{4{\pi}uKE{R^2}} \  ,\ 
\Pi = -\frac{R}{2h}\frac{\partial h}{\partial R}\epsilon \ .
\label{tang}
\end{equation}
These formulae show that the tangential stress vanishes whenever the function
$h$ does not  depend on $R$. Therefore in this case the material is not
sustained by any stress and  as a consequence the line element~(\ref{m01})
reduces to the dust one (Lema\^itre-Tolman-Bondi spacetime).  Since, in
all formulae,  $E$ and $h$ appear only in ratio, the value of $h$ on an
arbitrary curve $R=R_0(m)$ can be rescaled to unity, so that, 
in particular, the dust solutions can be characterized by $h=1$.
The function $R_0(m)$ plays the same role as played by
the initial mass distribution in the ``standard" comoving coordinates
(the inverse transformation from mass label to lagrangian label
being $r=R_0(m)$) while the function $E(m)$ plays the role of 
``energy function'' familiar from dust models (see~\cite{Magli02} for details).

The above recalled structure shows that a solution with tangential stresses
is identified (obviously modulo gauge transformations)
by a triplet of functions $\{g,E,h\}$.
This ``parameterization'' contains the dust spacetimes
as the ``subspace'' $\{g,E,1\}$.
In this way we can construct in a mathematically precise way,
a comparison between dust and tangential stress
evolutions by comparing the end state of the dust collapse 
($\{g,E,1\}$ with chosen $g$ and $E$) with the
endstates of the tangential stress solutions $\{g,E,h\}$
with the same $g$ and $E$ and different choices of the equation of state
$h$ (see~\cite{Magli02, JM1} for details).

The physical singularities of the spacetime described by the 
metric (\ref{m01}) correspond to infinite energy density 
and are given by $R=0$ or $K = 0$. 
At $R=0$ the shells of matter ``crush to zero size'' 
(shell-focusing singularities) while vanishing of $K$ implies intersection of 
different shells of matter (shell-crossing singularities).
We are interested here only in the central shell-focusing singularity.
It is important to distinguish the central singularity
(i.e. that shell focusing singularity occurring at $R=m=0$) from
the non-central singularities (singularities occurring at $R=0,m\neq 0$)
which are always covered in spacetimes with vanishing radial stresses,
as a simple consequence of mass conservation.

At $R=m=0$ the apparent horizon and the singularity form simultaneously.
Therefore, to analyse the causal nature of the singularity,
one has to study outgoing radial null geodesics
and check if there are some which meet the singularity in the past.
This can be done~\cite{Dwivedi} by
defining a quantity $x=R/2m^{\alpha}$ (where $1/3\leq \alpha\leq 1$)
and searching 
for a real positive solution $x_0$ of the following ``root equation''
\begin{eqnarray} \label{root}
x_0  &=&
\lim_{\stackrel{m\rightarrow0}{\scriptscriptstyle R\rightarrow0}}
\frac{m^{\frac{3}{2}(1-\alpha)}}{2\alpha}
\left[ -\frac{ R_{0,m} h(m,R_0)}{E(m) u(m,R_0)} 
+ \int^{2m^{\alpha}x}_{R_0}G dR \right] 
\left(\sqrt{(-1+\frac{E^2}{h^2})
m^{\alpha-1}+\frac{1}{x}} \right) \times 
\nonumber \\ &\ & \times
 \left(\frac{E}{h}  -\sqrt{-1+
\frac{E^2}{h^2} + \frac{m^{1-\alpha}}{x}} \right) .
\end{eqnarray}
This equation depends both on the choice of the equation of state $h$
and on the initial data. Setting $h=1$ one can deduce from it 
the final fate of the dust gravitational collapse~\cite{Jhingan97, Singh}, 
so that it can be used to study the way in which the end state of collapse for 
fixed initial data is altered by a non-trivial $h$. 

To analyse the roots equation, one faces with the fact that 
the integral appearing in formulae (\ref{root}) introduces a sort
of non-locality which is difficult to handle. As a consequence, it is very 
difficult to work in full generality and only a few cases have been analysed so
far. Among different cases with tangential stresses, a special role is
played by a system of counter-rotating particles (The Einstein Cluster). 
This system has several attractive features from the physical point of view 
since it mimics in a simple way the effect of rotation (see e.g.~\cite{Komer}).
So far, the only known results on the nature of the singularities forming in 
this model have been obtained in the HIN paper (such results will be recalled 
in the next Section).
 
It has been shown in paper~\cite{Magli01} that the Einstein cluster model
can be obtained from the general exact solution with tangential stresses by 
choosing a specific form of $h$:
\begin{equation}
h(m,R) = h_0(m)\left[1 + \frac{L^2(m)}{R^2} \right]^{1/2} .
\label{eqnstate01}
\end{equation}
In the above formula, the functions $h_0$ and $L$ are arbitrary. 
The function $L$ is the specific angular momentum of the particles
while only one 
among the functions $E$ and $h_0$ is independent. Therefore the
function $h_0$ can be used to normalize the value of $h$ to unity
on the initial data surface $R=R_0(m)$.
This choice does not affect the generality 
of the equation of state (\ref{eqnstate01}) within the Einstein cluster class
until the function $E$ is chosen. 

The final fate of the collapse turns out to depend on the behaviour 
of the physical quantities near origin, on the initial data surface.
Three such quantities are independent, and we shall use from now on  
$F$, $L^2$ and $f=E^2-1$. We assume these functions to be Taylor-expandable 
with respect to $r$ (the ``standard'' radial coordinate $r=R_0$) near $r=0$.
It is now easy to check (see~\cite{Magli01}) that regularity 
at the initial surface requires the following kind of expansions:
\begin{eqnarray} \label{exps}
F(r) &=& F_0r^3+F_nr^{n+3}+...
\\ 
f(r) &=& f_lr^l+...
\\
L^2(r) &=& L_kr^k+...
\end{eqnarray}
where $l\geq 2$, $k\geq 4$ and dots denote higher orders terms.
The $n$-term in $F(r)$ has been put in evidence 
because it characterizes the 
first non-vanishing derivative of the initial 
density profile; a positive $F_0$ and a negative $F_n$
are required for physical reasonability.

In the above formulae we allow for terms {\it of any parity}.
It is very important to specify carefully 
the meaning of this choice,
since a considerable
debate on this point has already taken place in the case of dust
(i.e. $L^2=0$ here). 

In the dust case, pioneering numerical results by Eardley and 
Smarr~\cite{Eardley}  were first investigated
analytically by Christodoulou~\cite{DChristodoulou} 
and Newman~\cite{Newman}. Both these authors required
the density and metric functions to be smooth ($C^{\infty}$)
with respect to a system of {\it cartesian} coordinates
near the centre (a function of this kind is sometimes called
extended analytic). Obviously, this implies 
that a power series expansion near $r=0$ can 
contain only {\it even} powers 
(for instance, if a linear term is absent but a cubic term
present in the mass $F(r)$, 
the initial density $\epsilon(r,0)=F'/4\pi r^2$ 
is only $C^{2}$ near the centre w.r. 
to a local cartesian system).
From the physical point of view, however, this restriction is 
unnecessary 
and actually rules out many interesting situations~\cite{JhinganSingh} .
Indeed, the results by Christodoulou and Newman 
were extended to the case of an arbitrary 
parity by Singh and Joshi~\cite{Singh} 
showing analytically, for instance, the existence of a transition
parameter at $n=3$ for 
marginally bound data (this structure will be very briefly recalled
in the next section).

Returning to the case of Einstein cluster, the extended analytic case 
here corresponds to even values of $k$ in the expansion of $L^2$ and was first 
considered in the HIN paper, while we allow here for odd values of $k$ as well 
as for even ones.

The expansions can now be translated 
in mass--area coordinates. In particular, we put
\begin{equation}
L^2(m) = \beta_k m^{k/3} +...\ , 
\label{l2}
\end{equation}
where $\beta_k$ is a positive constant.

\section{The endstate of collapse}

The dust models admit no globally regular solutions.
This is not true for general matter fields even within spherical symmetry.
For example, in homogeneous perfect fluid models, we can have globally
regular solutions which satisfy suitable energy conditions and
have regular initial data (see e.g.~\cite{Bonnor}); in other words, the 
endstate of a non-static model can be an eternally oscillating solution 
or a bounce-back scenario. It is, therefore, important to understand the 
qualitative behaviour of the motion as a preliminary step
before investigating the causal structure.

The equation of motion~(\ref{j}) governing the collapsing shells
can be conveniently
written as
\begin{equation}
u^2=\frac{Z(R,R_0)}{R(R^2+L^2)} \ ,
\label{motion}
\end{equation}
where the ``effective potential'' $Z$ is defined by
\begin{equation}
Z(R,R_0) = fR^3 +2mR^2 + L^2 (2m-R)\ .
\end{equation}
This function should be interpreted as an analogue of the Newtonian effective
potential governing the motion of the {\it fixed} shell $R_0$, so that
the allowed regions of the motion correspond to $Z\geq0$.
It is immediately seen that 
the sign of $f=E^2h_0^2-1$ 
governs the behaviour at large values of $R$, since
unboundedly large values of $R$ are permitted only for $f\geq0$.
The region near $R=0$ is, instead, always allowed for any $R_0\neq 0$
since $Z(0,R_0)=2mL^2>0$.
To investigate on the formation of the central singularity
(i.e. the singularity forming at $R=R_0=0$), let
$R=\zeta(t, R_0) R_0$.
To the lowest order in $R_0$ 
one has 
\begin{equation}
Z \approx  (
- \beta_k \zeta {R_0}^{k-4} + 2 F_0\zeta^2 +f_0 \zeta^3 + 2\beta_k F_0
{R_0}^{k-2} + \beta_k \zeta^3 {R_0}^{k-4} + \beta_k f_0
\zeta^3 {R_0}^{k-2})R_0^5 \ ,
\label{allowed}
\end{equation}
(recall that 
$F=F_0R_0^3$, $f=f_0R_0^2$ and $L^2 = \beta_k R_0^k$ to the lowest order).
Near $\zeta=0$, the above equation 
gives $Z\approx -\beta_4 \zeta R_0^5< 0$ for  
$k=4$, and therefore 
the singularity does not form.
Since it is the positivity of $\beta_4$ which does not allow 
singularity formation, tangential stresses regularize 
the corresponding ($\beta_4 = 0$) dust solutions. 
For $k>4$ one has 
$Z\approx 2F_0\zeta^2 R_0^5>0$ and the singularity always forms 
(the case $k=4$ was firstly studied by Evans~\cite{ev}, 
while the HIN paper contains the analytic case of even $k$).

For $k>4$ we have to analyse further the dynamics of the non-central shells,
which depends on the existence of roots of the cubic equation $Z=0$.
Due to the above remarks on the behaviour of $Z$ for big
and small values of $R$, we can {\it a priori} expect
the following kinds of situations.
If $R_1$, $R_2$ and $R_3$ are the roots of the cubic
and $R_1 < R_2 < R_3$, one has the following sets of allowed regions
\label{centre}
\begin{eqnarray}
0 \leq R \leq R_1, & R_2 \leq R \leq R_3 , \, & f < 0 , \nonumber \\ 
0 \leq R \leq R_1, & R_2 \leq R, \, & f \geq 0 ,
\label{allow}
\end{eqnarray}
where, of course, the forbidden regions between $R_2$ and $R_3$
can disappear if either one or both of these two roots vanish or are not
positive (thereby giving 
$0\leq R\leq R_1$ for $f<0$ and $R\geq 0$ for $f>0$, respectively).

The fine details are quite involved and were given for the first time by 
Bondi~\cite{Bondi}. The point of view here, however, is slightly different since
we are interested only in the dynamics of the shells near the central one
to check whether the central singularity becomes trapped or not.
A simple way to do this is to evaluate the effective potential on curves of 
the type $R=\lambda F$ with $\lambda>2$. For the region near centre we obtain
\begin{equation}
Z (\lambda F, R_0) \approx (2-\lambda)\beta_k F_0 R_0^{k+3} + 2
\lambda^2 F_0^3 R_0^9  .
\label{Approx}
\end{equation}
For $k \leq 5$ the first term is the leading one and it is negative.
For $k=6$ it is possible to find values $\lambda F$
such that $Z(\lambda F, R_0)$ is negative
if the quadratic equation $(2-\lambda) \beta_6 F_0 + 2 F_0^3 \lambda^2=0$
has a positive root. This in turn implies that
the quantity $D:=\sqrt{\beta_6}/F_0$ must be greater than four (HIN).
In both these cases ($k=5$ and $k=6$ for $D>4$) the central singularity is 
naked. Indeed, $\lambda F$ is not an allowed value since $Z$ is negative at it.
But $F$ goes to zero as $R_0^3$, and therefore the shells near the centre
must have initial data $R=R_0$ greater than $2F$.
For sufficiently small $R_0$, this will hold true also for 
$\lambda F$, so that a region of the kind $R>R_2>2F$ containing $R=R_0$
must exist. But the corresponding dynamics can never cross the forbidden region 
to reach values smaller than $2F$. Therefore, the region near the centre is 
untrapped {\it eternally} in these cases: the apparent horizon
does not form and therefore the singularity is naked.
This phenomenon has been firstly discovered in the HIN paper for 
the case $k=6, D > 4$. Interestingly enough, this kind of eternal singularity 
is completely different from the naked singularities arising in dust 
spacetimes, which are usually visible only for a finite amount of time before 
apparent horizon formation. A peculiar characteristic of the singularities 
here is that they are not accompanied by non-central singularities,
since all the shells near the central one remain regular eternally
(the region near $R=0$ is forbidden for them). 

For $k=6, D\leq 4$ as well as for any $k>6$ the above argument cannot be 
applied. The special case $k=6, D=4, f=0$ can, however, be treated
analytically since the integral appearing in formula (\ref{m10})
can be expressed in terms of arbitrary functions. The singularity in this 
case turns out to be always naked; again independently from the
choice of the initial matter distribution (HIN). 

To understand the nature of the singularities in the remaining possible cases,
we have to resort to the root equation. The integral which appears in 
formula~(\ref{root}) cannot be calculated explicitly, however, the existence 
of a naked singularity can be probed as follows (for simplicity we 
shall consider the case $E(m) =1$).
We expand the integrand near centre in the powers of $m$ and express the
integral as the sum over successive integrated terms of the series. 
The proof that this is possible makes use of Lebesgue's dominated convergence
theorem, as shown in details in the Appendix. 

For $k = 6$ equation~(\ref{thatsit}) gives 
\begin{equation}
x_0  = \lim_{\stackrel{m\rightarrow0}{\scriptscriptstyle R\rightarrow0}}
\frac{1}{2\alpha} \left[ 
\frac{nF_n}{9{\sqrt{2}}}\frac{m^{(2n-9\alpha+3)/6}}{F_0^{(2n+9)/6}}
- \frac{2^{3/2}\beta_6}{15 F_0^{1/6}} m^{(7-9\alpha)/6} -  \frac{2}{3}x^{3/2} +
\cdots \right] \left(\frac{m^{1-\alpha}}{x}- \frac{1}{\sqrt{x}} \right) \ ,
\label{FROOT}
\end{equation}
where dots stand for terms going as ${\cal O}(m^{\frac{n-1}{3}})$.   
In this equation it is possible to fix uniquely the value of $\alpha$ 
for each choice of $n$, in such a way that the resulting algebraic equation 
in $x_0$ has a positive root (see Table 1). Therefore the singularity is always
naked for $k=6$, independently from the choice of the initial density 
distribution. 

As recalled in the previous section, the value of the root
gives the tangent of the escaping geodesics near the singularity.
It is somewhat unclear if a physical content can be given
to the actual (``numerical'') value of this tangent, although it
is very likely that such an interpretation could be found in the near future.
In any case, we can get some insight into the effect of the state equation
on naked singularity formation looking {\it qualitatively} to the role of the 
parameter $\beta_k$ in the roots. With this aim, the values
of the roots for the various possible initial distributions of mass 
in presence of a non-zero $\beta_6$ (counter-rotating particles)
can be compared with the corresponding values for dust ($\beta_6 =0$)
(see Table 1).

\begin{table}[hbtp]
\centering
\begin{tabular}{|l|l|c|c||c|c|} \hline
 & Initial data & \multicolumn{4}{c|}{root $(x_0)^{3/2}$}\\
\cline{3-6}
 & $F_0R_0^3+F_nR_0^{n+3}$  & Einstein cluster & singularity  & dust 
& singularity \\ \hline
1 & n=1 & $\displaystyle{\frac{-F_1}{2^{5/2}F_0^{11/6}}}$ & visible
&  $\displaystyle{\frac{-F_1}{2^{5/2}F_0^{11/6}}}$ &visible \\ \hline
2 & n=2 &
$\displaystyle{\frac{6{\beta_6}F_0^2 - 5F_2}{20{\sqrt{2}}F_0^{13/6}}}$
& visible
& $\displaystyle{\frac{- F_2}{4{\sqrt{2}}F_0^{13/6}}}$
& visible \\
\hline
3 & n=3 & &
& roots for & \ 
\\ 
& & $\displaystyle{\frac{3\beta_6}{10{\sqrt{2}}F_0^{1/6}} }$
& visible & $\xi < \xi_c$ & transition \\ \hline
4  & n $>$ 3 & $\displaystyle{\frac{3\beta_6}{10{\sqrt{2}}F_0^{1/6}} }$
& visible
& - & black hole \\ \hline
\end{tabular}
\caption{The endstate of counter-rotating particles for $k=6$ vs dust }
\end{table}

At $n=1$ the inhomogeneity dominates over the effect of counter-rotation:
both the endstates are naked singularities and the roots coincide. 
At $n=2$ a contribution coming from the counter-rotation appears and
both the endstates are still naked singularities. At $n=3$ the root equation 
for dust spacetimes becomes a quartic which has positive roots 
only if the a-dimensional quantity $\xi :=F_3/(2^{3/2}F_0^{5/2})$
is less than  $\xi_c =-(26+15\sqrt{3})/2$~\cite{Singh}. Therefore, $\xi_c$ can 
be called critical parameter: at $\xi =\xi_c$ a transition occurs and
the endstate of collapse turns from a naked singularity to a black hole.
In the case of the Einstein cluster instead the root becomes independent
from $n$ at $n=3$ and is always present for $\beta_6 >0$ making the 
endstate always a naked singularity. For $n>3$ dust always forms a black hole 
while here the endstate is always a naked singularity.

The existence of such ``tenacious'' naked singularities which are not sensible 
to changes in the initial mass distribution, first discovered in the HIN paper,
raises the question if these singularities are already present in dust models 
or are peculiar of spacetimes with stresses. Actually it turns out that such 
special dust spacetimes exist, as we are going to show.

Observe that the energy function of the HIN model 
is of the form $f(r)=f_4r^4+...$ (where $f_4=-2^4F_0^2$), i.e., 
the expansion of this function starts from the quartic terms.
Now one can ask for the final fate of those {\it dust}
spacetimes which have initial velocity profiles of this kind.
To the best of our knowledge, the answer to this question is not present in the 
literature since all previous studies on dust spacetimes have assumed a 
non-vanishing second order term.
The causal structure can, however, be analysed easily
and it turns out that such solutions actually have the same 
causal structure of HIN: the singularity is always visible~\cite{notaroot}.

It remains to consider the cases $k\geq 7$. For $k=7$ formula~(\ref{thatsit}) 
gives 
\begin{equation}
x_0  = \lim_{\stackrel{m\rightarrow0}{\scriptscriptstyle R\rightarrow0}}
\frac{1}{2\alpha} \left[
\frac{nF_n}{9{\sqrt{2}}}\frac{m^{(2n-9\alpha+3)/6}}{F_0^{(2n+9)/6}}
- \frac{2^{1/2}\beta_7}{5F_0^{1/6}} m^{3(1 -\alpha)/2} -  
\frac{2}{3}x^{3/2} +
\cdots \right] \left(\frac{m^{1-\alpha}}{x} - \frac{1}{\sqrt{x}} \right) \ ,
\label{finroot}
\end{equation}
where dots stand for terms going to zero as ${\cal O}(m^{\frac{k+n-7}{3}})$.
This equation leads to a structure of the endstates
which is similar to that occurring in dust (second column of table one)
for $ n= 1$ and $2$.
At the transition 
($n=3$), the region of naked singularities formation is characterized 
by the inequality $\tilde \xi<\xi_c$ where 
$\tilde\xi = \xi - 3\beta_7/(5{\sqrt{2}} F_0^{1/6})$ and 
$\xi$ is the dust parameter recalled above.
Since $\beta_7$ is positive, a sector of the black hole region in dust 
is uncovered by the effect of counter-rotation; for $n > 3$ 
it remains uncovered for strong  enough 
rotation parameter ($\beta_7/F_0^{1/6} > -5\sqrt{2}\xi_c/3$).

Finally, for $k> 7$ the causal structure of the singularity is identical to 
that of marginally bound Lema\^itre-Tolman-Bondi spacetimes, since all the 
terms coming from rotation vanish near the singularity (a similar situation 
has been recently found to occur in a completely different model with 
tangential stresses~\cite{Barve}).

\section{Discussion}

In recent years, a big effort has been undertaken to understand the final fate
of the gravitational collapse of a dust cloud. Indeed, one can safely assert 
that this end state is now completely known in dependence of the choices of the
initial data.

The dust model is the most simple model which can be conceived in gravitational
collapse and in the absence of general proofs (and indeed even of mathematical 
formulations of) of cosmic censorship theorems, it was meant to be used as a 
tool to get insights in to more general collapse situations. We are, therefore,
in some 
sense ready to approach the problem with general stresses and indeed some 
analytical results are beginning to be known. In this respect our full 
understanding of dust spacetimes should hopefully be used as a starting point.

As a simple generalization of the dust models, we proposed to analyse 
spacetimes with tangential stresses and constructed a tool for doing that using
mass--area coordinates. The first results on the nature of the singularities 
for such spacetimes were obtained in the HIN paper for the case of the Einstein
cluster model. In the present paper we have 
extended the HIN results to the non-extended analytic case.

The picture which arises is very intriguing. The key role is played by the 
parameter $k$ measuring the ``strength'' of angular momentum near the centre 
($L^2\approx \beta_k R_0^k$ as $R_0$ tends to zero). For $k=4$ the solution is 
regular and tangential stresses avoid singularity formation. For $k=5$ we have 
central singularity formation, as in dust, but the neighbouring shells are 
regular; all such shells are untrapped forever
allowing the central singularity to be eternally visible. 
This phenomenon is clearly connected to singularity theorems 
since all the solutions considered here satisfy the energy conditions
and therefore, in the presence of trapping the shells, would have to become 
singular. For $k=6$ a sort of transition takes place, since
for $D=\sqrt{\beta_6}/F_0 \leq 4$ also non-central shells can become singular,
the exact HIN solution playing the role of separation point. 
For $k=7$ the remaining effect of counter-rotation becomes
very weak and it is effective only at the $n=3$ transition 
from black holes to naked singularities, shifting the 
transition parameter $\xi_c$ and uncovering part of dust black hole  
region. The naked singularities 
continue to exist for strong enough rotation parameter $\beta_7$.
For $k > 7$ the final fate is the same as in dust models.  

From the physical point of view, this picture is due to the non-convexity of 
the state function which acts as a ``source of acceleration''
near the centre avoiding the formation of apparent horizon and
therefore enforcing nakedness.
The structure of the spectrum of endstates
thus depends in a peculiar way on the fact that
the Einstein cluster model, when interpreted as a 
material characterized by a macroscopic  equation of state,
does not admit a local minimum of the energy density.
This is readily seen since 
the derivative of the state function $h$ with respect to $R$
is strictly negative and proportional to $-L^2$ near the centre,
while equations of state  which have to be expected for, say,
nuclear matter at high densities should admit a status of (local)
minimization of internal energy.
However, the HIN results, as well as our results here,
do raise the interesting question
if the effects of rotation might become
dominant with respect to those of a non-trivial equation of state 
at high densities; when the hypothesis of average geodesic motion
(which is the basis of the relative simplicity of the counter-rotating particles
model) is lost. To approach this question one should turn to the 
(so far unsolved) difficulties of the 
non-spherical collapse scenario.

Remaining in the safety zone of spherical symmetry, our results here
support the view that 
formation of naked singularities 
has to be expected in {\it {generic}} situations of spherical 
collapse~\cite{JD, CMP}.

\section*{Acknowledgements}

Interesting discussions with Pankaj Joshi are gratefully acknowledged.

\section*{Appendix}

Consider the function
\begin{equation}
I_{1}(m;\alpha):=\int_{R_0}^{2m^{\alpha}x} G(m,R)dR
\label{test}\tag{A1}
\end{equation}
where $G(m,R)$ is defined by equation~(\ref{m11}). Due to the
Mean value theorem
there exists
$\chi(m) \in (R_0,2m^{\alpha}x)$  such that
\begin{equation}
I_{1}(m;\alpha):= (2m^{\alpha}x -R_0) G(m,\chi(m)) .
\label{funda}\tag{A2}
\end{equation}
Since $R_0 \sim m^{1/3}$ as $m$ goes to zero, $1/3\leq \alpha \leq 1$
and $x$ is finite and positive, both $m^{-1/3}\chi (m)$ and $m^\alpha/\chi (m)$
are finite and positive as $m$ goes to zero.
Therefore we can evaluate the right hand
side of equation~(\ref{funda}) using~(\ref{m11}) as follows
\begin{align}
\int_{R_0}^{2m^{\alpha}x} G(m,R)  dR & = m^{-1}
\frac{(2m^{(3\alpha-1)/3}x - F_0^{-1/3} +\cdots )}{2\sqrt{2} 
(1+
\beta_k m^{(k-2)/3}F_0^{2/3})^{1/2}}\left(\frac{\chi(m)}{m^{1/3}}\right)^{1/2}
\left[ 1+ {\beta_k}m^{(k-6\alpha)/3}\left(\frac{m^{\alpha}}{\chi(m)}\right)^2 
\right.  \notag  \\ & \left. 
+ \frac{\beta_k}{2}m^{(k-4)/3}\left(\frac{\chi(m)}{m^{1/3}}\right)F_0^{2/3}
- \frac{\beta_k}{2}m^{(k-3-3\alpha)/3}\left(\frac{m^{\alpha}}{\chi(m)}\right)
+ \cdots \right]^{-3/2}
\notag \\ & \times
 \left[\left(1+ {\beta_k}m^{(k-6\alpha)/3}\left(\frac{m^{\alpha}}{\chi(m)}
\right)^2 \right)^2 + 
\frac{\beta_k}{6}(k-2)m^{(k-4)/3}\left(\frac{\chi(m)}{m^{1/3}}\right)
F_0^{2/3} \right. \notag \\ & \left.
 - \frac{{\beta_k}^2}{3}m^{(2k-5-3\alpha)/3}
\left(\frac{m^{\alpha}}{\chi(m)}\right) F_0^{2/3}
- \frac{k \beta_k}{6}m^{(k-3-3\alpha)/3}\left(\frac{m^{\alpha}}{\chi(m)}\right)
+ \cdots \right] ,
\label{needed}\tag{A3}
\end{align}
where dots stand for terms of
higher order in positive powers of $m^{n/3}$. 
For $k \geq 6$ only the factor $m^{-1}$ is divergent,
all the other terms being finite due to $\alpha \leq 1$. 
Therefore, the limit
\begin{equation}
I_{2}(m;\alpha):= \lim_{m\rightarrow0}
\int_{R_0}^{2m^{\alpha}x} m \  G(m,R)dR ,
\label{life}\tag{A4}
\end{equation}
is convergent. Hence,
using Lebesgue's dominated convergence theorem, we can expand the
integrand near the centre ($m = 0$) in the 
leading powers of $m$ and integrate successive terms. This gives
\begin{gather}
I_{2}(m;\alpha) =
\lim_{m\rightarrow0}
\left[\frac{-1}{3\sqrt{2}F_0^{1/2}} +
\frac{F_n m^{n/3}}{6\sqrt{2}F_0^{(2n+9)/6}}
+ \frac{\beta_k\sqrt{2}(k-4)}{15F_0^{1/6}}m^{(k-4)/3}   ~ ~ ~
\right. \notag \\
\left.
 -\frac{2}{3}m^{(3\alpha-1)/2}x_0^{3/2} + \frac{\beta_k(2k-9)}{6}m^{(2k
-9 + 3\alpha)/2}x_0^{1/2} + {\cal O}(m^{(k-4+n)/3}) \right] ,
\label{intexp}\tag{A5}
\end{gather}
Using this equation in~(\ref{root}),
we can write the root equation in the form (valid for $k\geq 6$)
\begin{align}
x_0  = &
\lim_{\stackrel{m\rightarrow0}{\scriptscriptstyle R\rightarrow0}}
\frac{m^{\frac{3}{2}(1-\alpha)}}{2\alpha}
\left[ \frac{nF_n m^{(n-3)/3} }{9{\sqrt{2}}F_0^{(2n+9)/6}}
- \frac{\beta_k {\sqrt{2}}(k-4)}{15F_0^{1/6}} m^{(k-7)/3} -
\frac{2}{3}m^{3(\alpha-1)/2}{x_0}^{3/2} + \right.
\notag \\  &
 + {\cal O}(m^{(k-7+n)/3)})\left]
( \frac{m^{1-\alpha}}{x}- \frac{1}{\sqrt{x}} \right. ) .
\label{thatsit}\tag{A6}
\end{align}

\vfill\eject


\begin{thebibliography}{99}
\bibitem{notaroot}
Using the root equation a simple calculation shows 
that the values of the roots are the following:
$-F_1/(2^{5/2}F_0^{11/6})$ for $n=1$,
$(24F_0^2 - 5F_2)/(20 \sqrt{2} F_0^{13/6})$ for $n=2$
and $6/(5{\sqrt{2}}F_0^{1/6})$ for any $n\geq 3$.
\end{thebibliography}
\end{document}